\documentclass[
]{ceurart}

\usepackage{listings}
\usepackage{graphicx}
\usepackage{grffile}
\usepackage{url}
\usepackage{booktabs}
\usepackage{multirow}
\usepackage{acronym}
\acrodef{vqa}[VQA]{Visual Question Answering}

\begin{document}

\copyrightyear{2022}
\copyrightclause{Copyright for this paper by its authors.
  Use permitted under Creative Commons License Attribution 4.0
  International (CC BY 4.0).}

% \conference{Woodstock'22: Symposium on the irreproducible science,
%   June 07--11, 2022, Woodstock, NY}
\conference{ISWC'22: The 21st International Semantic Web Conference, October 23--27, 2022, Hangzhou, China}

\title{VGStore: A Multimodal Extension to SPARQL for Querying RDF Scene Graph}

% \tnotetext[1]{You can use this document as the template for preparing your
%   publication. We recommend using the latest version of the ceurart style.}

\author[1]{Yanzeng Li}[%
email=liyanzeng@stu.pku.edu.cn
]
\address[1]{Wangxuan Institute of Computer Technology (WICT), Peking University, China}

\author[2]{Zilong Zheng}[%
email=zlzheng@bigai.ai
]
\address[2]{Beijing Institute for General Artificial Intelligence (BIGAI), Beijing, China}

\author[2]{Wenjuan Han}[%
email=hanwenjuan@bigai.ai
]

\author[1]{Lei Zou}[%
email=zoulei@pku.edu.cn
]\cormark[1]

\cortext[1]{Corresponding author.}

\begin{abstract}
  Semantic Web technology has successfully facilitated many RDF models with rich data representation methods. It also has the potential ability to represent and store multimodal knowledge bases such as multimodal scene graphs. However, most existing query languages, especially SPARQL, barely explore the implicit multimodal relationships like semantic similarity, spatial relations, etc. We first explored this issue by organizing a large-scale scene graph dataset, namely Visual Genome, in the RDF graph database. Based on the proposed RDF-stored multimodal scene graph, we extended SPARQL queries to answer questions containing relational reasoning about color, spatial, etc. Further demo (i.e., \texttt{VGStore}) shows the effectiveness of customized queries and displaying multimodal data.
\end{abstract}


\maketitle

\section{Introduction}

% In recent years, multimodality has become a popular research topic in natural language processing and computer vision.
Over the recent years, we have witnessed an explosive growing trend on multimodal models due to the increasing computing power and massive multimodal datasets.
% Multimodal models have gained a lot of popularity in recent years due to the breakthroughs and increasing potential applications in both computer vision (CV) and natural language processing (NLP). 
% While different modailities like vision and language are characterized by different properties, they complement and enhance each other.
Despite of inspiring performance that keeps updating on various multimodal benchmarks, the interpretability and reasonability have recently been challenged by researchers, namely, models are memorizing multimodal statistical features rather than understanding the joint information among them. For example, \ac{vqa}, a representative multimodal testbed for vision understanding, requires model to reason over images and answer questions. However, the current mainstream models still depend on end-to-end training by fitting input signals to ground truth answers, while neglecting the underlying visual relations and semantics.
% The most representative Multimodal (MM) application is Visual Question Answering (VQA), which requires cross-modality interactions to answer the question. The current mainstream end-to-end VQA models depend on supervised fitting to annotated datasets by large deep learning models, which lack interpretability and reasonability, and are difficult to transfer to a new domain or adapt to the open domain. 
% These issues have been deeply studied in Knowledge Base Question Answering (KBQA), which aims to answer Natural Language Questions~(NLQ) by querying external Knowledge Base (KB).

To address these issues, we leverage an intrinsically explainable task, Knowledge Base Question Answering (KBQA), which aims to answer Natural Language Questions by referring to external Knowledge Base (KB).
Semantic Parsing (SP)-based KBQA is a mainstream technique to solve the such QA problem via parsing a question into a KB query (such as SPARQL)~\cite{zhang-etal-2022-crake}.
% , it has strong advantages in reasonability and interpretability
The latest works are devoted to improving the performance of natural language understanding and parsing, while neglecting the expressiveness of KB queries, limiting the application of SP-based methods in multimodal datasets.
Although RDF has sufficient representation capability to describe multimodal data, the lack of multimodal semantic relationships in standard SPARQL has become a major challenge in applying SP-based KBQA methods to multimodal domains. Researchers have attempted to extend SPARQL for this purpose. For example, SPARQL-MM~\cite{kurz2014sparql} proposed to use custom aggregation functions to access media fragments. However, previous works still suffer from limitations in extensibility and vague semantic representation, resulting in rare applications. 
Specifically, the custom aggregation functions introduced in SPARQL-MM are easy to understand by human beings but hard to understand and be expressed by the machine. This is because it brings significantly extra complexity to the query statement, e.g., if multimodal query statements in SPARQL-MM are used as query conditions, it inevitably leads to the union or nested query; however, the SP-based methods only support simple queries in the foreseeable future.

In this demo, we designed an ontology to organize the multimodal scene graph and store it with RDF. Furthermore, we implemented semantic multimodal SPARQL queries by extending the SPARQL engine, enabling the ability to answer questions related to multimodal information such as visual and spatial reasoning.

\section{Storing Visual Genome with RDF}

Visual Genome (VG)~\cite{krishna2017visual} is a large-scale dataset for fine-grained scene graphs, with rich annotations of images, regions, objects, as well as their relations\footnote{\url{http://visualgenome.org/VGViz/explore} demonstrates the dataset.}. A synset from WordNet~\cite{miller1995wordnet} is introduced to link different scene graphs via the lexical relations between literals of the object relations. In addition, VG provided 1,445,332 relevant questions for 108,077 images, which are difficult to be answered by traditional SP-based KBQA methods because the SPARQL engine does not support the arithmetic opeartions needed to answer these multimodal questions.
\begin{figure}[t]
\centering
\includegraphics[width=\linewidth]{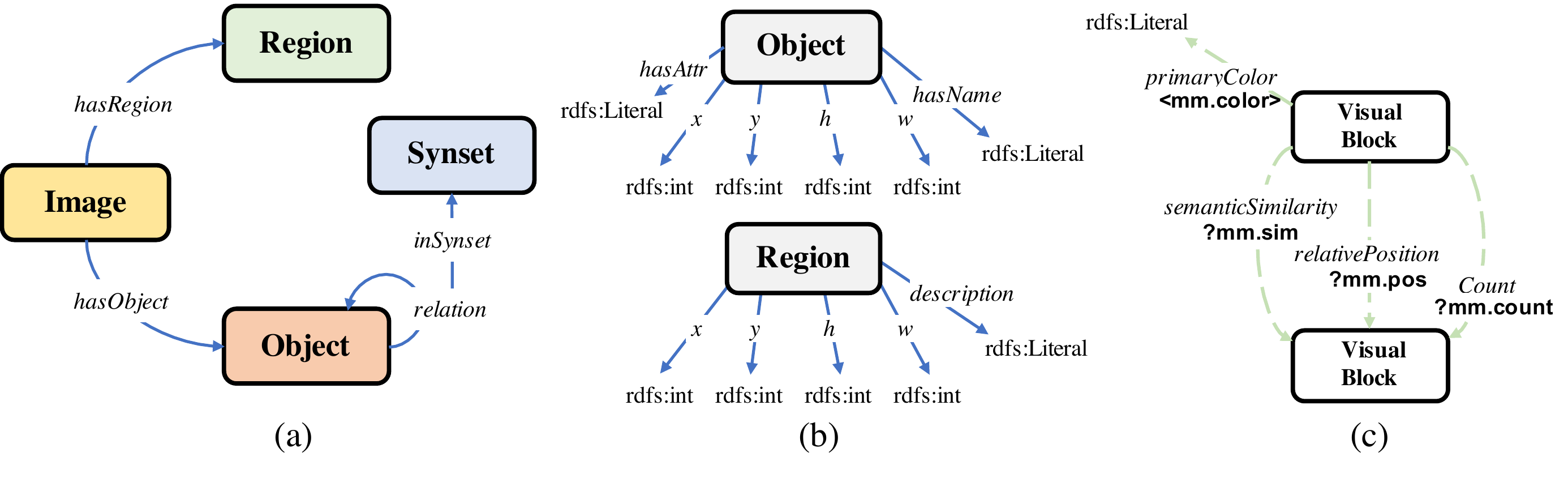}
\caption{(a) The example of relations between classes in the ontology of visual genome. (b) The example of data properties of the class. (c) The implicit semantic relations between visual blocks.} 
\label{ontology}
\end{figure}

For querying convenience, we formalize the elements of VG in RDF. Fig. \ref{ontology}(a) shows the designed ontology of RDF-stored VG (RDF-VG)\footnote{Due to space limitations, the detail of data processing and ontology organization are attached to the code repository.}. Fig. \ref{ontology}(b) demonstrates properties of the defined classes in RDF-VG. The properties $(x,y,h,w)$ determine the visual block of region or object by tailoring image. We store RDF-VG in \texttt{gStore}~\cite{zeng2018redesign}, which is a graph-oriented RDF data management system supporting complex SPARQL queries on graph data. % 

\section{Querying Multimodal Information via SPARQL}

\begin{figure}[t]
\centering
\includegraphics[width=\linewidth, trim=0 30 0 0,clip]{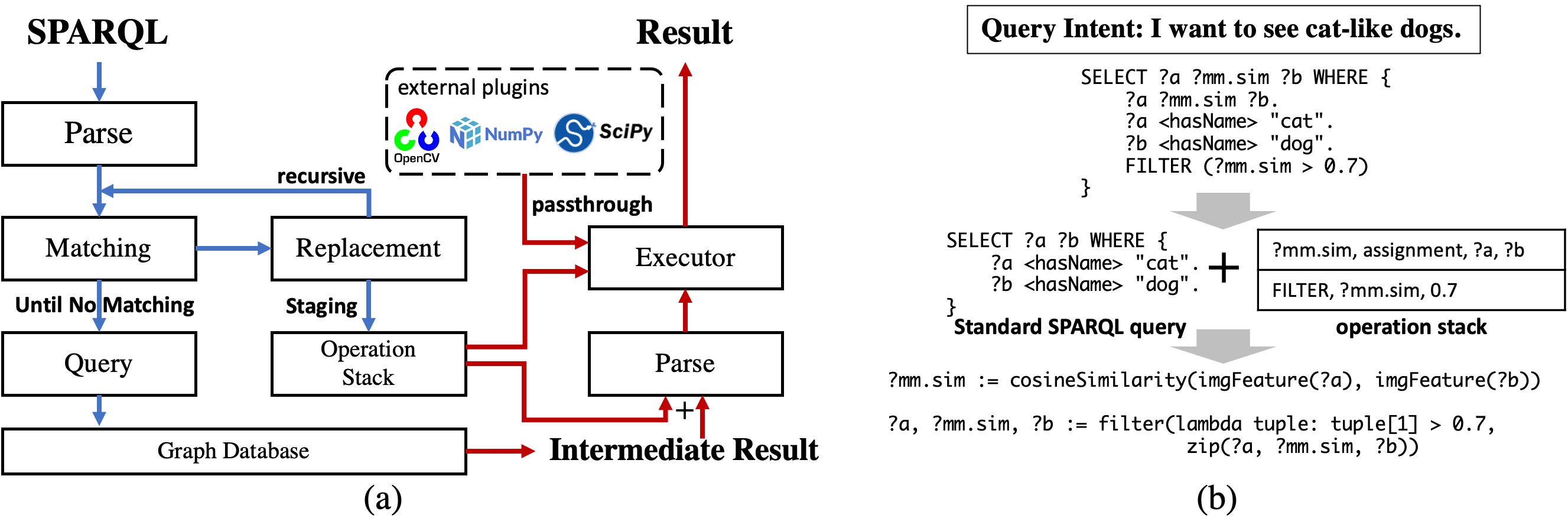}
\caption{Left: The diagram of \texttt{VGStore}. Right: An example of \texttt{VGStore} handling the multimodal SPARQL query (Python-like pseudocode).} % ImageFeature, e.g. cv2.calcHist
\label{sparql}
\end{figure}

Traditional SPARQL engines (such as our backbone - \texttt{gStore}) cannot perform queries involving multimodality and thus cannot directly answer such questions (e.g. what color is this cat?). Therefore, we developed a \texttt{VGStore} extension based on standard SPARQL grammar and \textit{pyparsing}~\cite{mcguire2007getting} to parse custom predicates (as Fig. \ref{ontology}(c) shows) for arbitrary query patterns, enabling the ability of traditional graph databases to perform multimodal queries by passing through the extra computing requirements to third-party tools (e.g. OpenCV, Torch, etc.). The architecture of \texttt{VGStore} is shown in Fig. \ref{sparql} (Left).
\begin{table}[b]
\centering
% \small
\caption{Part of the supported non-standard querying clauses. The {\ttfamily ?a} and {\ttfamily ?b} indicate the regular query variables.}\label{tab1}
\begin{tabular}{p{0.35\linewidth}p{0.59\linewidth}}
\toprule 
\multicolumn{1}{l}{\textbf{Customized Triple Pattern}} & \multicolumn{1}{c}{\textbf{Description}} \\
\midrule
{\ttfamily ?a {\bf{?mm.sim}} ?b.} & Represent the semantic similarity between {\ttfamily ?a} and {\ttfamily ?b}.  \\
{\ttfamily ?a {\bf $\langle${mm.color}$\rangle$} ?color.} & Output the primary color of {\ttfamily ?a} to variable {\ttfamily ?color}. \\
{\ttfamily ?a {\bf{?mm.pos}} ?b.} & Represent the relative position relation between {\ttfamily ?a} and {\ttfamily ?b}.  \\
{\ttfamily ?a {\bf{?mm.count}} ?b.} & Count the number of component {\ttfamily ?b} in image {\ttfamily ?a}.  \\
\midrule
{\multirow{2}{*}{\ttfamily FILTER(...)}} & Filter the results via customized variables, \newline e.g., {\ttfamily FILTER(?mm.sim > 0.5).
}  \\\midrule
{\multirow{2}{*}{\ttfamily ORDER BY ...}} & Sort the results via customized variables, \newline e.g., {\ttfamily ORDER BY DESC(?mm.count).
}  \\
\bottomrule
\end{tabular}
\end{table}

\texttt{VGStore} analyzes, matches, and replaces the clauses in the original SPARQL query that contain custom predicates. It recursively replaces all non-standard query patterns with the standard SPARQL syntax, and stores the replacement process in an operation stack temporarily.
The standard SPARQL query can be handed over to the backbone graph database for execution. After getting the result, the inverse operation is successively performed according to the staged replacement operations in the stack. Finally, the intent of the original SPARQL query would be restored.
Fig. \ref{sparql} (Right) demonstrated how a multimodal query containing the non-standard custom predicate variable is to be executed. 
Table 2 illustrates part of the supported query patterns in \texttt{VGStore}, covering questions involving color, counting, and relative position in the VG question-answer dataset, which account for 15.0\%, 11.4\%, and 7.0\% of all questions, respectively. 
Other simple questions (e.g., ``What is this?'') can be expressed and queried by native SPARQL directly, and the remaining non-factual questions (e.g., ``What is this man's motivation?'') or inference questions (e.g., ``When was the picture taken?'') are out of scope in this demonstration.

\section{Discussion and Next Step}

Although \texttt{VGStore} successfully supports multimodal queries by extending virtual predicates, it still has some limitations. 
\texttt{VGStore} is written in Python, which brings additional latency in runtime, and it is possible to reduce the performance loss by native support in the SPARQL engine.
In addition, when the \texttt{VGStore} queries large data, and there are multiple extended query statements, it would cause severe performance problems. This drives us to schedule and parallelize the third-party tool calls. 

\texttt{VGStore} currently only supports several basic query patterns (as listed in Table \ref{tab1}) specialized for the RDF-VG dataset, and does not adapt to other VQA datasets, nor does it support richer query patterns. 
Therefore, our next steps for improvements include supporting more query patterns and extending the applicability of \texttt{VGStore} to more graph databases with large-scale multimodal graphs.

\begin{figure}[t]
\centering
\includegraphics[width=\linewidth, trim=0 20 0 0,clip]{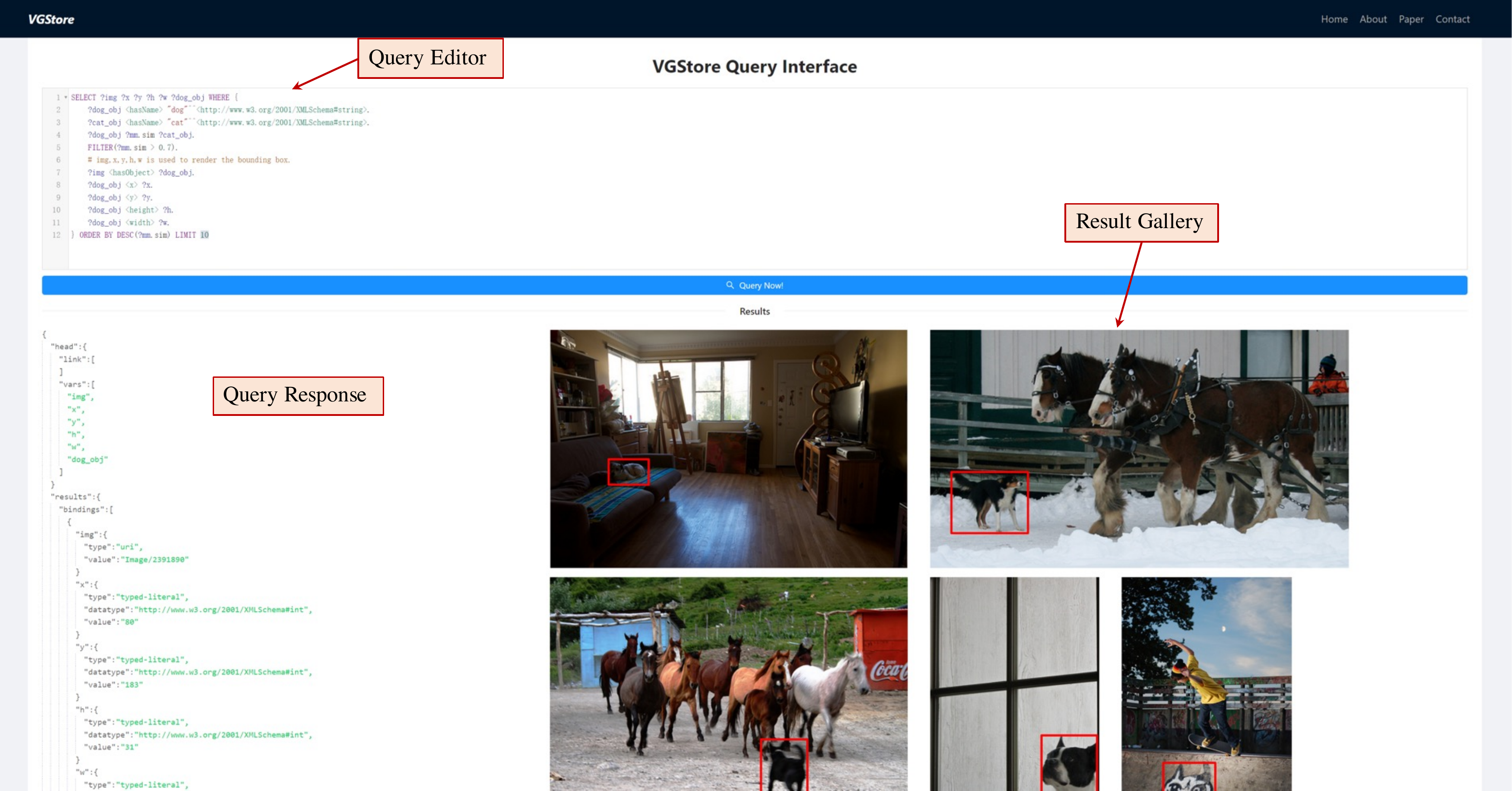}
\caption{The interface in use for querying the RDF-stored visual genome scene graph with customized query pattern.} 
\label{screenshot}
\end{figure}

\section{Demonstration}
% The demo will showcase the web user interface of \texttt{VGStore} for querying multimodal SPARQL on RDF-VG, as Fig. \ref{screenshot} shows. 
% This paper presented the demo of \texttt{VGStore}, an extension to SPARQL for querying multimodal information on RDF. With this demo, the solution to the main challenge of SP-based multimodal KBQA laid the foundation. 
This paper presented \texttt{VGStore}, an extension to SPARQL for querying multimodal information on RDF. The demo showcased the web user interface of \texttt{VGStore} for querying multimodal SPARQL on RDF-VG, as shown in Fig. \ref{screenshot}. With this demo, we provided a potential solution to the main challenge of SP-based multimodal KBQA and laid the foundation of multimodal knowledge graph. 
Our code of RDF-VG builder, preliminary parser and frontend of demonstration are available at \url{https://github.com/pkumod/VGStore}.

\begin{acknowledgments}
  This work was supported by NSFC under grant 61932001, U20A20174. The corresponding author of this paper is Lei Zou (zoulei@pku.edu.cn). We sincerely thank reviewers for their valuable comments and advises.
\end{acknowledgments}

%%
%% Define the bibliography file to be used
\bibliography{sample-ceur}

\begin{thebibliography}{6}
\expandafter\ifx\csname natexlab\endcsname\relax\def\natexlab#1{#1}\fi
\providecommand{\url}[1]{\texttt{#1}}
\providecommand{\href}[2]{#2}
\providecommand{\path}[1]{#1}
\providecommand{\DOIprefix}{doi:}
\providecommand{\ArXivprefix}{arXiv:}
\providecommand{\URLprefix}{URL: }
\providecommand{\Pubmedprefix}{pmid:}
\providecommand{\doi}[1]{\href{http://dx.doi.org/#1}{\path{#1}}}
\providecommand{\Pubmed}[1]{\href{pmid:#1}{\path{#1}}}
\providecommand{\bibinfo}[2]{#2}
\ifx\xfnm\relax \def\xfnm[#1]{\unskip,\space#1}\fi
%Type = Inproceedings
\bibitem[{Zhang et~al.(2022)Zhang, Zhang, Li, and Zou}]{zhang-etal-2022-crake}
\bibinfo{author}{M.~Zhang}, \bibinfo{author}{R.~Zhang},
  \bibinfo{author}{Y.~Li}, \bibinfo{author}{L.~Zou},
\newblock \bibinfo{title}{Crake: Causal-enhanced table-filler for question
  answering over large scale knowledge base},
\newblock in: \bibinfo{booktitle}{Findings of the Association for Computational
  Linguistics: NAACL 2022}, \bibinfo{year}{2022}, pp.
  \bibinfo{pages}{1787--1798}.
%Type = Inproceedings
\bibitem[{Kurz et~al.(2014)Kurz, Schaffert, Schlegel, Stegmaier, and
  Kosch}]{kurz2014sparql}
\bibinfo{author}{T.~Kurz}, \bibinfo{author}{S.~Schaffert},
  \bibinfo{author}{K.~Schlegel}, \bibinfo{author}{F.~Stegmaier},
  \bibinfo{author}{H.~Kosch},
\newblock \bibinfo{title}{Sparql-mm-extending sparql to media fragments},
\newblock in: \bibinfo{booktitle}{European Semantic Web Conference},
  \bibinfo{organization}{Springer}, \bibinfo{year}{2014}, pp.
  \bibinfo{pages}{236--240}.
%Type = Article
\bibitem[{Krishna et~al.(2017)Krishna, Zhu, Groth, Johnson, Hata, Kravitz,
  Chen, Kalantidis, Li, Shamma et~al.}]{krishna2017visual}
\bibinfo{author}{R.~Krishna}, \bibinfo{author}{Y.~Zhu},
  \bibinfo{author}{O.~Groth}, \bibinfo{author}{J.~Johnson},
  \bibinfo{author}{K.~Hata}, \bibinfo{author}{J.~Kravitz},
  \bibinfo{author}{S.~Chen}, \bibinfo{author}{Y.~Kalantidis},
  \bibinfo{author}{L.-J. Li}, \bibinfo{author}{D.~A. Shamma}, et~al.,
\newblock \bibinfo{title}{Visual genome: Connecting language and vision using
  crowdsourced dense image annotations},
\newblock \bibinfo{journal}{International journal of computer vision}
  \bibinfo{volume}{123} (\bibinfo{year}{2017}) \bibinfo{pages}{32--73}.
%Type = Article
\bibitem[{Miller(1995)}]{miller1995wordnet}
\bibinfo{author}{G.~A. Miller},
\newblock \bibinfo{title}{Wordnet: a lexical database for english},
\newblock \bibinfo{journal}{Communications of the ACM} \bibinfo{volume}{38}
  (\bibinfo{year}{1995}) \bibinfo{pages}{39--41}.
%Type = Article
\bibitem[{Zeng and Zou(2018)}]{zeng2018redesign}
\bibinfo{author}{L.~Zeng}, \bibinfo{author}{L.~Zou},
\newblock \bibinfo{title}{Redesign of the gstore system},
\newblock \bibinfo{journal}{Frontiers of Computer science} \bibinfo{volume}{12}
  (\bibinfo{year}{2018}) \bibinfo{pages}{623--641}.
%Type = Book
\bibitem[{McGuire(2007)}]{mcguire2007getting}
\bibinfo{author}{P.~McGuire}, \bibinfo{title}{Getting started with pyparsing},
  \bibinfo{publisher}{" O'Reilly Media, Inc."}, \bibinfo{year}{2007}.

\end{thebibliography}

\end{document}